\documentclass[amsmath,amssymb,nofootinbib,prd]{revtex4}

\usepackage{graphicx}
\usepackage{epsfig}
\usepackage{rotate}
\usepackage{amsmath}
\usepackage{amssymb}
\usepackage{amsfonts}
\usepackage{color}

\def\hksqrt{\mathpalette\DHLhksqrt}
\def\DHLhksqrt#1#2{\setbox0=\hbox{$#1\sqrt{#2\,}$}\dimen0=\ht0
\advance\dimen0-0.2\ht0
\setbox2=\hbox{\vrule height\ht0 depth -\dimen0}%
{\box0\lower0.4pt\box2}}

\newcommand{\fnlt}{$f_{\rm NL}$ }
\newcommand{\sigfnl}{$\sigma(f_{\rm NL})$ }

\newcommand{\nngt}{$n_{\rm NG}$ }
\newcommand{\signng}{$\sigma(n_{\rm NG})$ }

\begin{document}

\title{Optimization of spectroscopic surveys for testing non-Gaussianity}

\author{Alvise Raccanelli$^{1,2}$, Olivier Dor\'{e}$^{1,2}$, Neal Dalal$^{3}$\\~}

\affiliation{
$^1$Jet Propulsion Laboratory, California Institute of Technology,
Pasadena CA 91109, USA \\
$^2$California Institute of Technology,
Pasadena CA 91125, USA \\
$^3$Department of Astronomy, University of Illinois, 1002 W.
Green St., Urbana, IL 61801, USA
}

\begin{abstract}
We investigate optimization strategies to measure primordial non-Gaussianity with future spectroscopic surveys. We forecast measurements coming from the 3D galaxy power spectrum and compute   constraints on primordial non-Gaussianity parameters $f_{\rm NL}$ and $n_{\rm NG}$. After studying the dependence on those parameters upon survey specifications such as redshift range, area, number  density, we assume a reference mock survey and investigate the trade-off between number density and area surveyed. We then define the observational requirements to reach the detection of $f_{\rm NL}$ of order 1. 
Our results show that while power spectrum constraints on non-Gaussianity from future spectroscopic surveys can be competitive with current CMB limits, measurements from higher-order statistics will be needed to reach a sub unity precision in the measurements of the non-Gaussianity parameter $f_{\rm NL}$.
\end{abstract}

\date{\today}

\maketitle


\section{Introduction}

A key part of our cosmological model is represented by the modeling of the physics of the primordial universe; the standard cosmological model includes cosmological inflation, a period of exponential expansion in the very early universe.
In this model, primordial cosmological perturbations are created from quantum fluctuations during the early period of accelerated expansion of the universe. Those perturbations formed the initial seeds for structure formation, and because of that the clustering properties of the matter distribution in the Universe on large scales allow us to probe the physics of that epoch.

Different models of cosmological inflation have different effects on the properties of the large-scale matter distribution and its evolution through cosmic time; one of the discriminants between different inflationary models is represented by the gaussianity (or deviation from it) of the probability distribution function of cosmological perturbations. For this reason, an  important goal for forthcoming cosmological experiments is to test whether initial conditions of the probability distribution function of cosmological perturbations deviate from gaussianity;
this can be done using the CMB (\cite{bartolo04, komatsu10} and references therein) or 
the large-scale structure of the Universe (see e.g.~\cite{matarrese00, dalal08, slosar08, matarrese2008, desjacques10, xia10ng, Raccanelliradio, Camera, Ferramacho, Ross13, RaccanelliISW14}).

Most recently, the Planck satellite measured \fnlt to be 2.7 $\pm$5.8 (68\% C.L.)~\cite{planckfnl}, thus constraining strongly the amount of primordial non-Gaussianity. 
However, since \fnlt of a few is expected just from the non-linear evolution of the gravitational potential (as e.g. the lensing-ISW contribution measured by Planck), even in the absence of hints of primordial non-Gaussianity, it remains important to measure the value of \fnlt as precisely as possible in the future, to check that our cosmological model and understanding of cosmological perturbations is correct.

In this paper, we assess the constraining power on non-Gaussianity of future spectroscopic surveys and investigate optimization strategies that will allow to exploit those capacities to the maximum. The paper is organized as follows. In Section~\ref{sec:ng} we introduce the observable effects of primordial non-Gaussianity, and in Section~\ref{sec:pk} we explain how we model the power spectrum in presence of non-Gaussianity. Our technique for obtaining the constraints is introduced in Section~\ref{sec:ff}. In Section~\ref{sec:survey} we present the parameters of the surveys we consider and study how the non-Gaussianity parameters depend on them. In Section~\ref{sec:opt} we investigate how one could optimize those surveys in order to be more effective in measuring non-Gaussianity parameters. Finally, in Section~\ref{sec:conclusions} we discuss our results and present our conclusions.


\section{Measuring non-Gaussianity with galaxy surveys}
\label{sec:ng}

The simplest models for inflation give rise to a nearly Gaussian distribution of the Bardeen potential $\Phi$ at primordial times. The most general expression of a deviation from Gaussianity at quadratic level can be expressed by a non-local relationship between the primordial Bardeen's potential $\Phi$ and a Gaussian auxiliary potential $\phi$.

Deviations from Gaussian initial conditions in the so called ``local type'' can be parameterized by the dimensionless parameter $f_{\rm NL}$ as~\cite{salopek, komatsu}:
\begin{equation}
\label{eq:fnl}
\Phi_{\rm NG}=\phi+f_{\rm NL}\left(\phi^2-\langle\phi^2\rangle\right) ,
\end{equation}
where $\Phi$ denotes Bardeen's gauge-invariant potential, which, on sub-Hubble scales reduces to the usual Newtonian peculiar gravitational potential.
Here $\phi$ is the Gaussian random field, and the second term, when $f_{\rm NL}$ differs from zero, gives the deviation from gaussianity;
in this paper we refer to the so-called ``local type" $f_{\rm NL}$ and we use the LSS convention (as opposed to the CMB one, where $f_{\rm NL}^{\rm LSS} \sim 1.3 f_{\rm NL}^{\rm CMB}$~\cite{xia10ng}).

In many scenarios the primary effect of the non-Gaussian correction results in a non-zero bispectrum, that can be written as:
\begin{equation}
\langle \Phi(k_1)\Phi(k_2)\Phi(k_3) \rangle = (2\pi)^3 \delta_D^3(k_1 + k_2 + k_3)B_\Phi(k_1, k_2, k_3) \, .
\end{equation}
In the local type, the bispectrum $B$ is:
\begin{equation}
B_\Phi(k_1, k_2, k_3) = f_{\rm NL} [2 P_\Phi(k_1)P_\Phi(k_2) + 2 \, \rm perm.] \, ,
\end{equation}
where:
\begin{equation}
\langle \Phi(k_1)\Phi(k_2) \rangle = (2\pi)^3 \delta_D^3(k_1 + k_2)P_\Phi(k_1, k_2) \, .
\end{equation}
As shown in e.g.~\cite{giannantonio11}, for 3D power spectrum analyses in galaxy surveys, the constraints on shapes different from the local type are considerably weakened; for this reason, in this work we focus only on the local type.

\subsection{Non-Gaussian bias}
\label{sec:ngbias}
A non-zero $f_{\rm NL}$ in Equation~\ref{eq:fnl} introduces a scale-dependent modification of the large-scale halo bias; we can write the total, non-Gaussian bias, as:
\begin{equation}
\label{eq:ng-bias}
b_{\rm NG}(z, k) = b_{\rm G}(z) + \Delta b(z, k)\, ; 
\end{equation}
the difference from the usual Gaussian bias, is (see e.g.~\cite{matarrese2008, dalal08, slosar08, desjacques10, xia10ng}): 
\begin{equation}
\label{eq:deltab}
\Delta b(z, k) = [b_{\rm G}(z)-1] f_{\rm NL}\delta_{\rm ec} \frac{3 \Omega_{0m}H_0^2}{c^2k^2T(k)D(z)}, 
\end{equation}
where $b_{\rm G}(z)$ is the usual bias calculated assuming gaussian initial conditions, assumed to be scale-independent, $D(z)$ is the linear growth factor and $\delta_{\rm ec}$ is the critical value of the matter overdensity for ellipsoidal collapse, $\delta_{\rm ec}=\delta_{\rm c}\hksqrt{q}$, with $q$ being a parameter fit with simulations~\cite{xia10ng}.
Given that we will base our calculations on a survey observing emission line galaxies, we use, for the gaussian part of the bias (see e.g.~\cite{pfs}):
\begin{equation}
\label{eq:gau-bias}
b_{\rm G}(z) = 0.9 + 0.4 z \, .
\end{equation}

\subsection{Running of $f_{\rm NL}$}

Some models of inflation, e.g. the ones with a variable speed of sound (such as Dirac-Born-Infeld models), have been shown to predict a scale-dependent non-Gaussian parameter \fnlt. A constraint on this scale-dependence would provide additional informations on the fundamental parameters of the underlying high-energy theories, such as the number of inflationary fields and their interactions (see e.g.~\cite{chen05, liguori06, khoury09, byrnes10, riotto11, kobayashi12}).

So far the only measurement of such running has been presented in~\cite{becker12} (which obtained an error of $\approx 0.15$); this represents an additional handle in probing inflationary physics, so it will be of great importance to set tight constraints on it. The running parameter \nngt is expected to be of the order of unity in most inflationary models. Its signatures in the CMB and LSS have been discussed in the literature (see e.g.~\cite{loverde08, sefusatti09, becker11}).

We consider the scenario that isolate the multi-field effects as in~\cite{shandera}, that gives, for the non-Gaussian part of the bias:

\begin{equation}
\Delta b_{\rm NG} (z, k, M) = f_{\rm NL}^{\rm eff} (M, n_{\rm NG}^{(m)}, k_*) \left(\frac{k}{k_*}\right)^{n_{\rm NG}^{(m)}}
\left[ [b_{\rm G}(z)-1] \delta_{\rm ec} \frac{3 \Omega_{0m}H_0^2}{c^2k^2T(k)D(z)} \right] \, ,
\end{equation}
where:
\begin{equation}
f_{\rm NL}^{\rm eff} (M, n_{\rm NG}^{(m)}, k_*) = \frac{\xi_m(k_p)}{2\pi^2\sigma(M)^2} \int_0^{\infty} k_1^2P_{\Phi}(k_1)M_{\rm R}^2(k_1) \left(\frac{k_1}{k_*}\right)^{n_{\rm NG}^{(m)}} \, .
\end{equation}
We then use the correction suggested in~\cite{desjacques1, desjacques2}, that gives a non-Gaussian, scale-dependent bias correction as:
\begin{equation}
\Delta b_{\rm NG} (z, k) = f_{\rm NL} k_*^{-n_{\rm NG}^{(m)}} \left( \frac{\sigma_{\alpha s}}{\sigma_{0s}} \right)^2 \left[ b_1 \delta_{\rm ec} +2 \left( \frac{\partial {\rm ln}\sigma_{\alpha s}}{\partial {\rm ln}\sigma_{0 s}}-1 \right) \right] 
\left[ [b_{\rm G}(z)-1] \delta_{\rm ec} \frac{3 \Omega_{0m}H_0^2}{c^2k^2T(k)D(z)} \right] \, ,
\end{equation}
where the spectral moments are defined as:
\begin{equation}
\sigma^2_{ns} = \int \frac{k^{2n}}{(2\pi)^3} P_\phi(k) \mathcal{M}_s^2(k) \, d^3 k \, .
\end{equation}
In this work we assume $k_*=0.04$; constraints for different values of such arbitrary fixed parameters can be easily converted when needed. 

\begin{center}
\begin{figure*}[tb!]
\includegraphics[width=0.49\columnwidth]{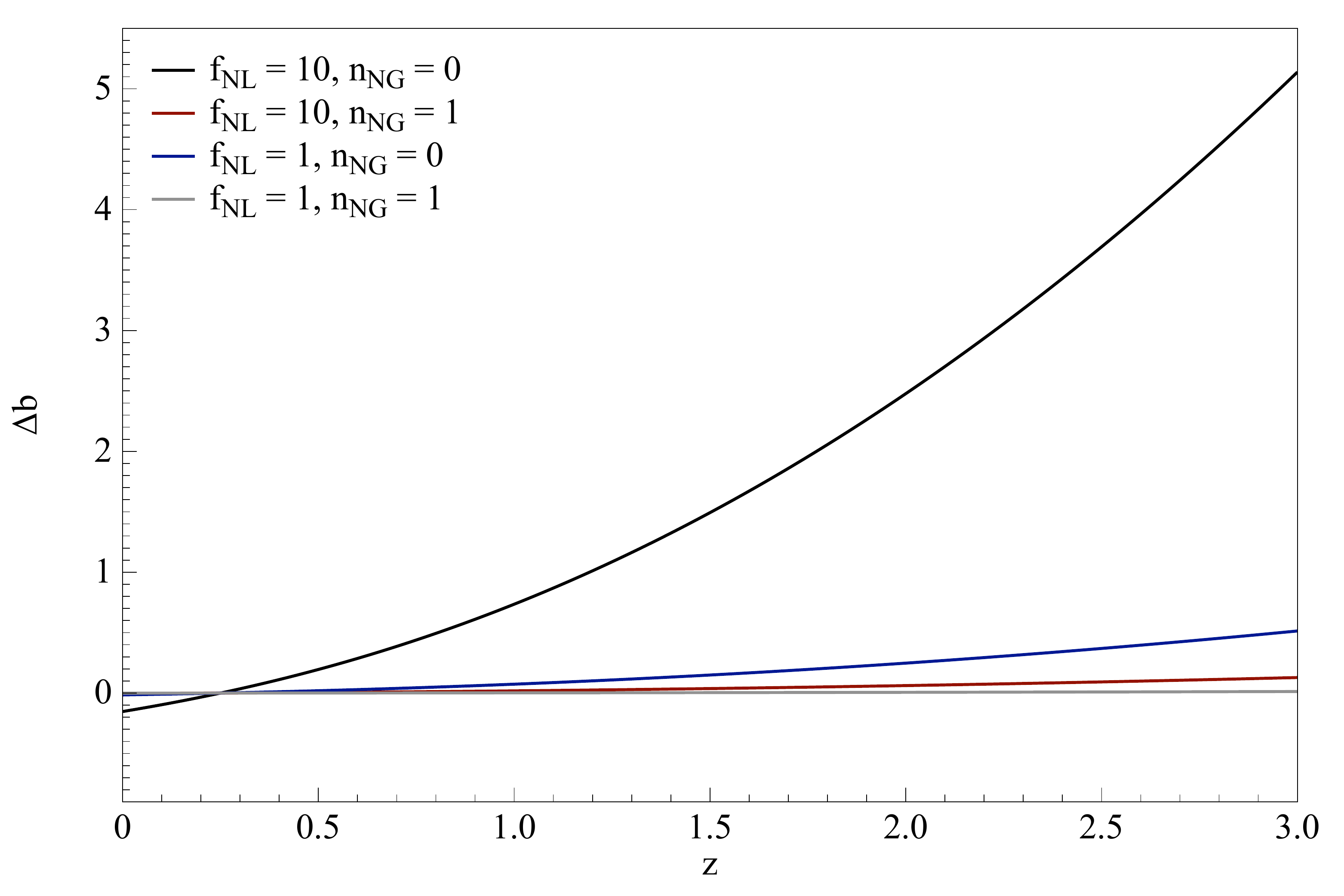}
\includegraphics[width=0.49\columnwidth]{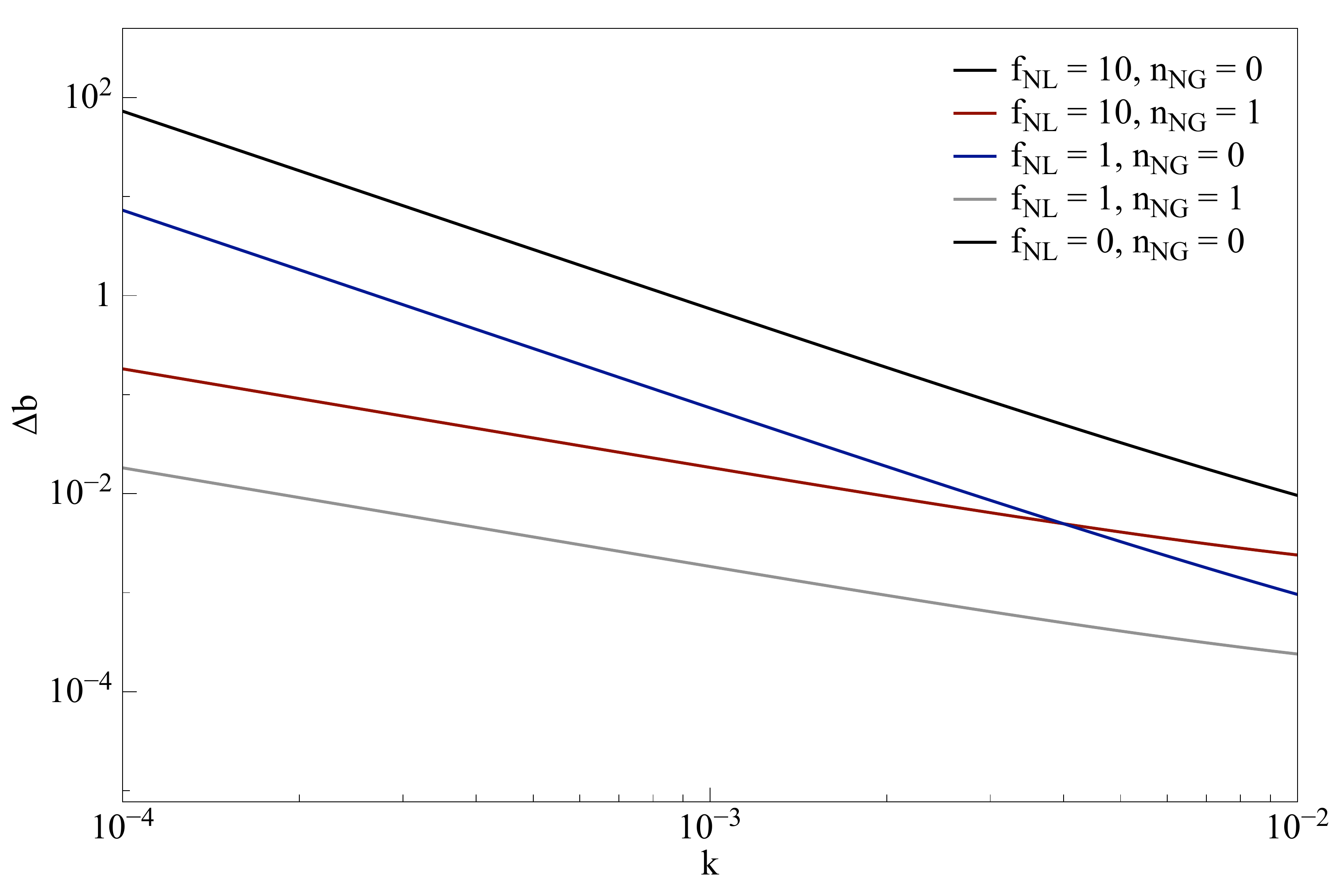}
\caption{Non-Gaussian corrections to the bias, for different combinations of $\{f_{NL}, n_{NG}\}$. {\it Left Panel}: as a function of redshift. {\it Right Panel}: as a function of scale.}
\label{fig:biasz}
\end{figure*}
\end{center}

\begin{center}
\begin{figure*}[htb!]
\includegraphics[width=0.87\columnwidth]{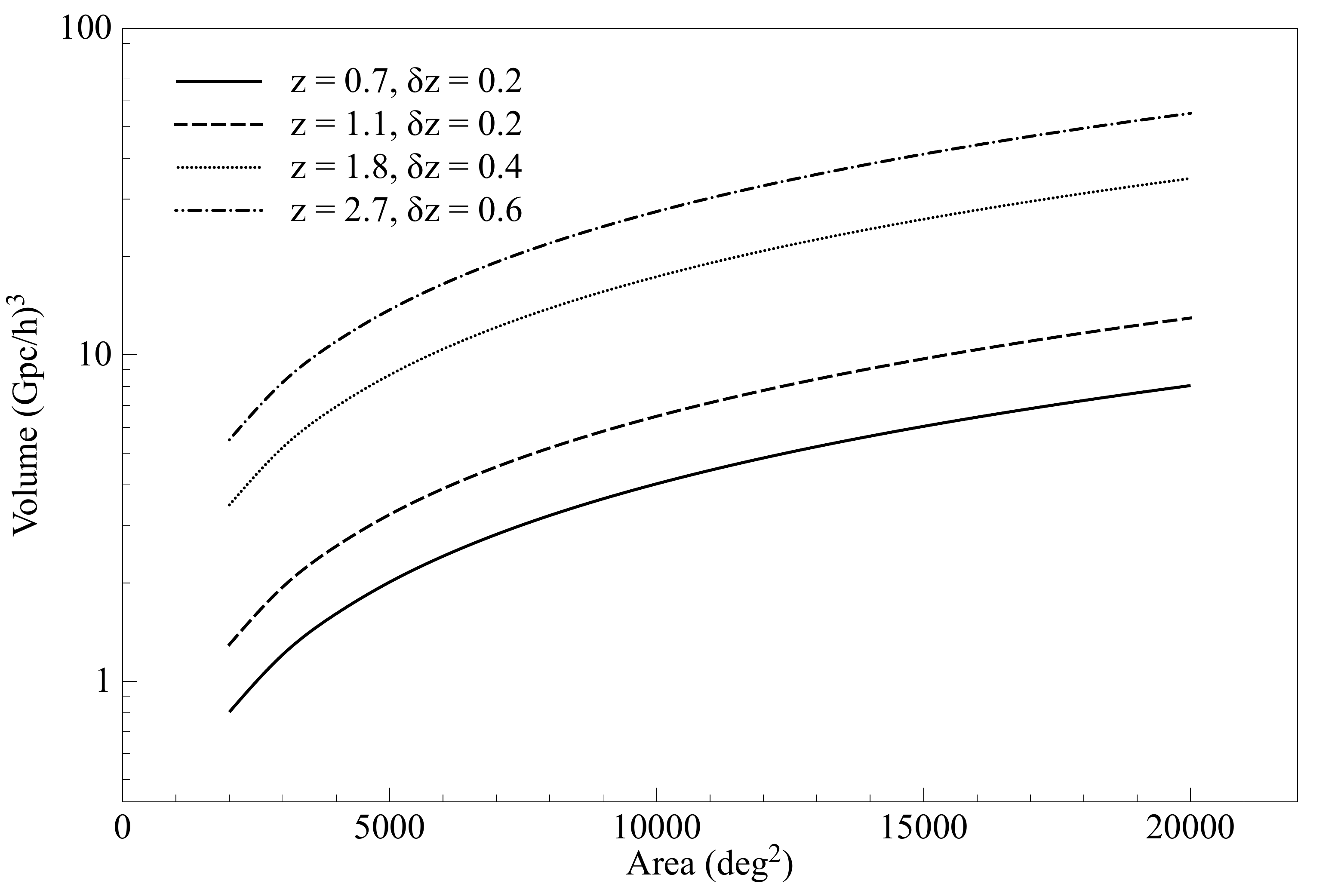}
\caption{
Cosmological volumes as a function of the area surveyed, for different redshift bins.
}
\label{fig:volumes}
\end{figure*}
\end{center}

From Figure~\ref{fig:biasz} it can be seen that the largest deviation is at higher redshift, and for larger scales. It is also interesting to notice that a positive $n_{NG}$ considerably reduces the effective modification due to a non-zero \fnlt, so that the deviation from the gaussian bias with $\{f_{NL}=1, n_{NG}=0\}$ is larger than the case $\{f_{NL}=10, n_{NG}=1\}$. In Figure~\ref{fig:volumes} we plot the cosmological volume probed at different redshifts and for different redshift bins; it is apparent how going to higher redshift will considerably increase the volume probed and so the possibility of constraining non-Gaussianity.

\section{Modeling the Power Spectrum}
\label{sec:pk}
The matter power spectrum depends on a variety of cosmological parameters, and for this reason its measurements has been used (together with its Fourier transform, the correlation function) to constraint e.g. dark energy parameters~\cite{samushia12}, models of gravity~\cite{Raccanelligrowth}, neutrino mass~\cite{dePutter12, Zhao12}, dark matter models~\cite{Cyr-Racine13, Dvorkin}, the growth of structures~\cite{Samushia13, Reid13}, and non-Gaussianity~\cite{Ross13}. 

We define the power spectrum as:
\begin{align}
P^s_g(k,\mu,z) = \left[ b_{\rm NG}(k,z) + f(z) \mu^2 \right]^2 P_m^r(k,z) + P_{shot}(z) \, ,
\end{align}
where the superscripts $r$ and $s$ indicate real and redshift-space, respectively, and the subscripts $m$ and $g$ stands for matter and galaxies; the shot noise contribution is taken to be:
\begin{align}
P_{shot}(z) = \frac{1}{\bar{n}_g(z)} \, ,
\end{align}
where the non-Gaussian bias is the one defined in Section~\ref{sec:ngbias}. The Redshift-Space Distortion (RSD) corrections come from the fact that the real-space position of a source in the radial direction in modified by peculiar velocities due to local overdensities; this effect can be modeled as~\cite{kaiser87, hamilton97}:
\begin{equation}
\label{eq:kaiser}
\delta^s(k) = \left( 1+\beta \mu^2 \right) \delta^r(k) \, ,
\end{equation}
where the parameter $f$ is defined as the logarithmic derivative of the growth factor:
\begin{equation}
f = \frac{d \, \rm ln \, D}{d \, \rm ln \, a},
\end{equation}
and in the $\Lambda$CDM model it can be parameterized as~\cite{linder05}:
\begin{align}
f(z) = \left[\Omega_m(z)\right]^\gamma.
\end{align}

In principle, additional effects could need to be taken into account when performing actual data analysis. Since we aim here at optimizing a survey for measuring non-Gaussianity, we will not take them into account but it should be noted that a precise analysis will require them. For example, Equation~\ref{eq:kaiser} assumes the plane-parallel approximation. When considering wide surveys and galaxy pairs with large angular separation, a more precise analysis involving wide-angle and GR corrections should be used (see e.g.~\cite{szalay97, matsubara99, szapudi04, papai08, raccanelli10rsd, yoo10a, samushia12, Bonvin11, Challinor11, yoo12, Jeong12, Bertacca12, Raccanelligrowth, Raccanelli13, Raccanelliradial}). However, including wide-angle corrections in the power spectrum requires a computationally very expensive treatment and it goes beyond the scope of this paper; their effects should not drastically modify our conclusions.
On very large scales, the modeling for the power spectrum needs to take in account General Relativity effects general relativistic corrections will be important (see e.g.~\cite{yoo10a, Bonvin11, Challinor11, yoo12, Jeong12, Bertacca12}).
When computing the power spectrum using a proper GR formalism, the corrections with respect to the newtonian case mimic a positive non-zero \fnlt (see e.g.~\cite{bruni}; this would be relevant only when \fnlt is very small~\cite{maartens12}).
However~\cite{Raccanelli13} showed that the two effects can be disentangled using their different angular dependence. A proper GR treatment would not change our conclusions, as we look at the precision in $\sigma_{f_{\rm NL}}$.
Nevertheless, when performing a real data measurement of a very large-scale survey it will be important to use a precise modeling of the large-scale power spectrum.

The observed radial distribution of sources is in redshift-space, so in principle the N(z) should be modified to take in account Redshift-Space Distortions~\cite{rassat09} (this can be modeled knowing the slope of the redshift distribution, the $\alpha$ parameter of~\cite{Raccanelliradio}); the radial distribution of sources is also modified by cosmic magnification~\cite{loverde07}. These two effects modify number of sources per bin (see also~\cite{Nock10}), but this is a second order effect (and we are using predicted radial distributions), so we will not include a proper treatment of that.
A careful treatment of all these corrections is left for future work.


\section{Fisher Forecast}
\label{sec:ff}
Given the specifications of a survey, the Fisher matrix analysis allows us to estimate the errors on the cosmological parameters around the fiducial values (see e.g. \cite{fisher35, tegmark97}). We write the Fisher Matrix for the power spectrum in the following way:
\begin{align}
\label{eq:FM}
F_{\alpha\beta} = \int_{z_{\rm min}}^{z_{\rm max}} dz \int_{k_{\rm min}}^{k_{\rm max}}dk  \int_{-1}^{+1}d\mu
& \left[\frac{\bar{n}_g(z) P(k,\mu,z)}{1+\bar{n}_g(z) P(k,\mu,z)}\right]^2
\frac{V_s(z) k^2}{8\pi^2 \left[P(k,\mu,z)\right]^2} \frac{\partial P(k,\mu,z)}{\partial \vartheta_\alpha}\frac{\partial P(k,\mu,z)}{\partial \vartheta_\beta} B_{nl} \, ,
\end{align}
where $\vartheta_{\alpha(\beta)}$ is the $\alpha(\beta)$-th cosmological parameter, $V_s$ is the volume of the survey and $\bar{n}_g$ is the mean comoving number density of galaxies.
The last term accounts for the non-linearities induced by the BAO peak~\cite{seo}:
\begin{equation}
B_{nl} = e^{-k^2\Sigma_{\perp}^2 -k^2 \mu^2 \left( \Sigma_{||}^2 -\Sigma_{\perp}^2 \right) } ,
\end{equation}
and $\Sigma_\bot=\Sigma_0D$, $\Sigma_{||}=\Sigma_0(1+f)D$, where $\Sigma_0$ is a constant
phenomenologically describing the nonlinear diffusion of the BAO peak due to
nonlinear evolution. From N-body simulations its numerical value is 12.4 $\rm h^{-1} Mpc$ and seems to depend linearly on $\sigma_8$, but only weakly on $k$ and cosmological parameters.

To perform the Fisher analysis, we parameterize our cosmology using: 
\begin{equation}
\label{eq:paratriz} 
{\bf P} \equiv \{ w_{\rm 0}, w_{\rm a}, n_{\rm s}, h, \sigma_8, b, \omega_{\rm m}, \omega_{\rm b}, \Omega_{\rm \Lambda}, \Omega_{\rm K}, n_{\rm g}, \gamma, f_{\rm NL}, n_{\rm NG} \} .
\end{equation}

While in a real analysis one should compute the entire Fisher Matrix of Equation~\ref{eq:FM}, we don't analyze the constraints coming from a full Fisher analysis, as non-Gaussianity parameters are weakly correlated with the other parameters usually analyzed in power spectrum analyses; moreover, in the full case the result will depend on the parameters included in the analysis, the cosmological model assumed, and over which parameters to marginalize.
In our case we limit ourselves to the calculation of $F_{\rm \alpha \beta}$, where $\{\alpha, \beta\} = \{ f_{\rm NL}, n_{\rm NG} \}$, and we assume fiducial values of $\{5, 0.1\}$, but the results do not depend strongly on this choice.


\section{Survey Parameters}
\label{sec:survey}
We want to study how \sigfnl depends on survey parameters; to do that, we begin assuming a fiducial survey with the parameters of Table~\ref{tab:sur}; for the redshift bins, we assume the same bins of the PFS experiment (see~\cite{pfs}), but we add a high-redshift bin of $2.4 \le z \le 3.0$. We assume a number density that gives $nP = 1.0$ at $k=0.1$ in each redshift bin. We then proceed to analyze where the best constraining power comes from in terms of redshift range, number density and area surveyed.

\begin{table}
\begin{center}
\begin{tabular}{ |p{3.33cm}|p{3.33cm}|}
	  \hline
	  \textbf{Parameter} & \textbf{Value - fiducial} \\
	  \hline
	  Field of view & 1.5 $\deg$\\
	  \hline
	  Total observing time & 800 hours \\
	  \hline
	  Exposure time & 15 mins  \\
	  \hline
	  Overhead time & 3 mins \\
	  \hline
	  $z_{\rm max}$ & 3.0\\
	  \hline
	  $nP (@ \, k=0.1)$ & 1.0\\
	  \hline
	\end{tabular}
\caption{Parameters for the reference survey considered.}
\label{tab:sur}
\end{center}
\end{table}

For our reference survey we define:
\begin{equation}
\label{eq:nights}
Area =  \frac{\tau}{n_v(t+o)} \times {\rm FoV} \, ,
\end{equation}
where $\tau$ is the total observing time, $n_v$ is the number of visits, $t$ is the exposure time, $o$ is the overhead of the instrument, and ${\rm, FoV}$ is its field of view.

\subsection{Redshift Range}
\begin{center}
\begin{figure*}[htb!]
\includegraphics[width=0.45\columnwidth]{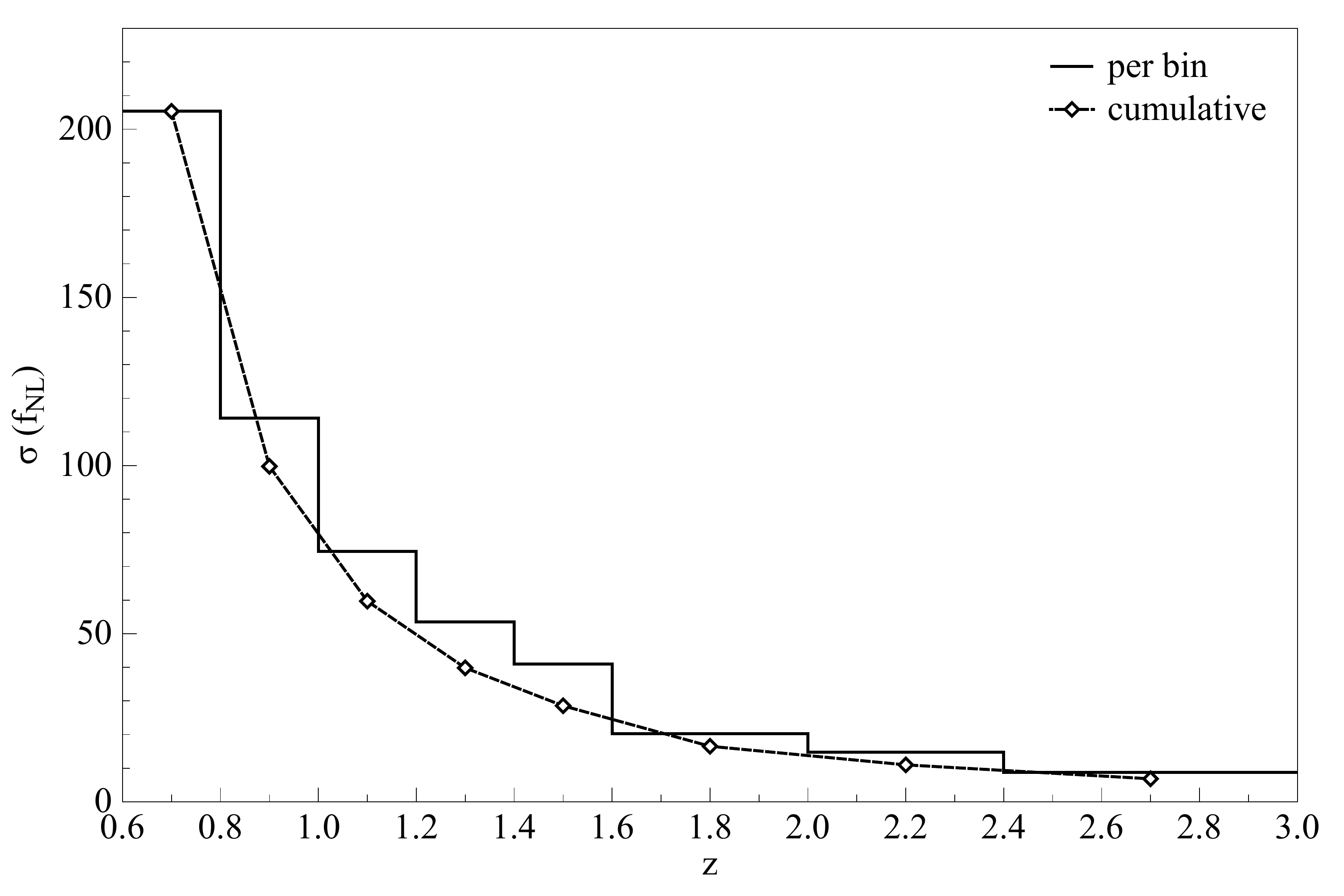}
\includegraphics[width=0.45\columnwidth]{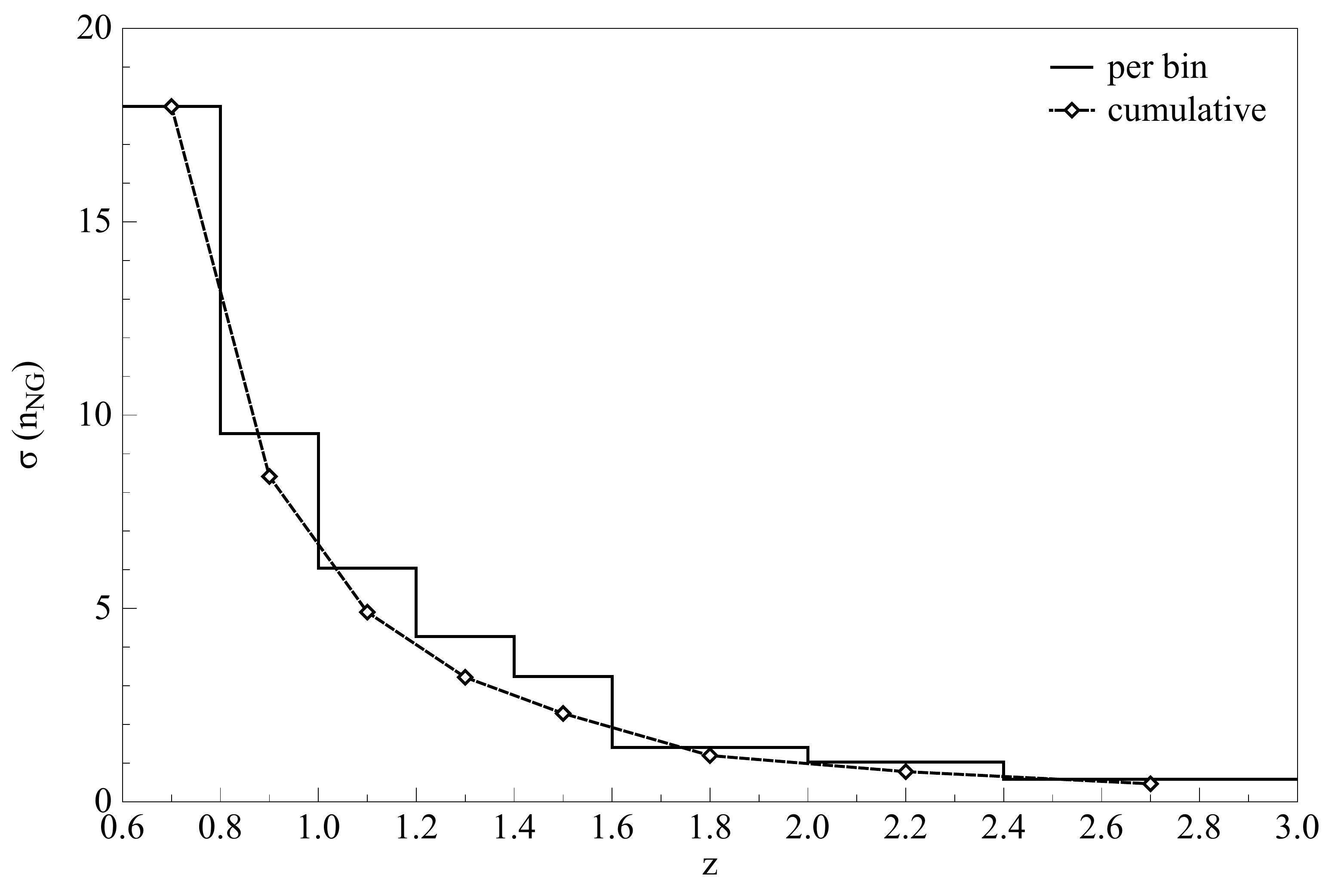}
\caption{Effect of adding z-bins. The solid line shows the constraining power of individual bins, the dashed line with symbols the cumulative effect of adding bins.}
\label{fig:fnl_z}
\end{figure*}
\end{center}

In this section we look at how the constraints on non-Gaussianity parameters depend on the redshift range and which z-bin has more constraining power.
In Figure~\ref{fig:fnl_z} we plot the constraining power for both \fnlt (left panel) and its running \nngt (right panel) for the redshift bins we consider, individually and when adding them together.

\subsection{Source Density}
\begin{center}
\begin{figure*}[htb!]
\includegraphics[width=0.45\columnwidth]{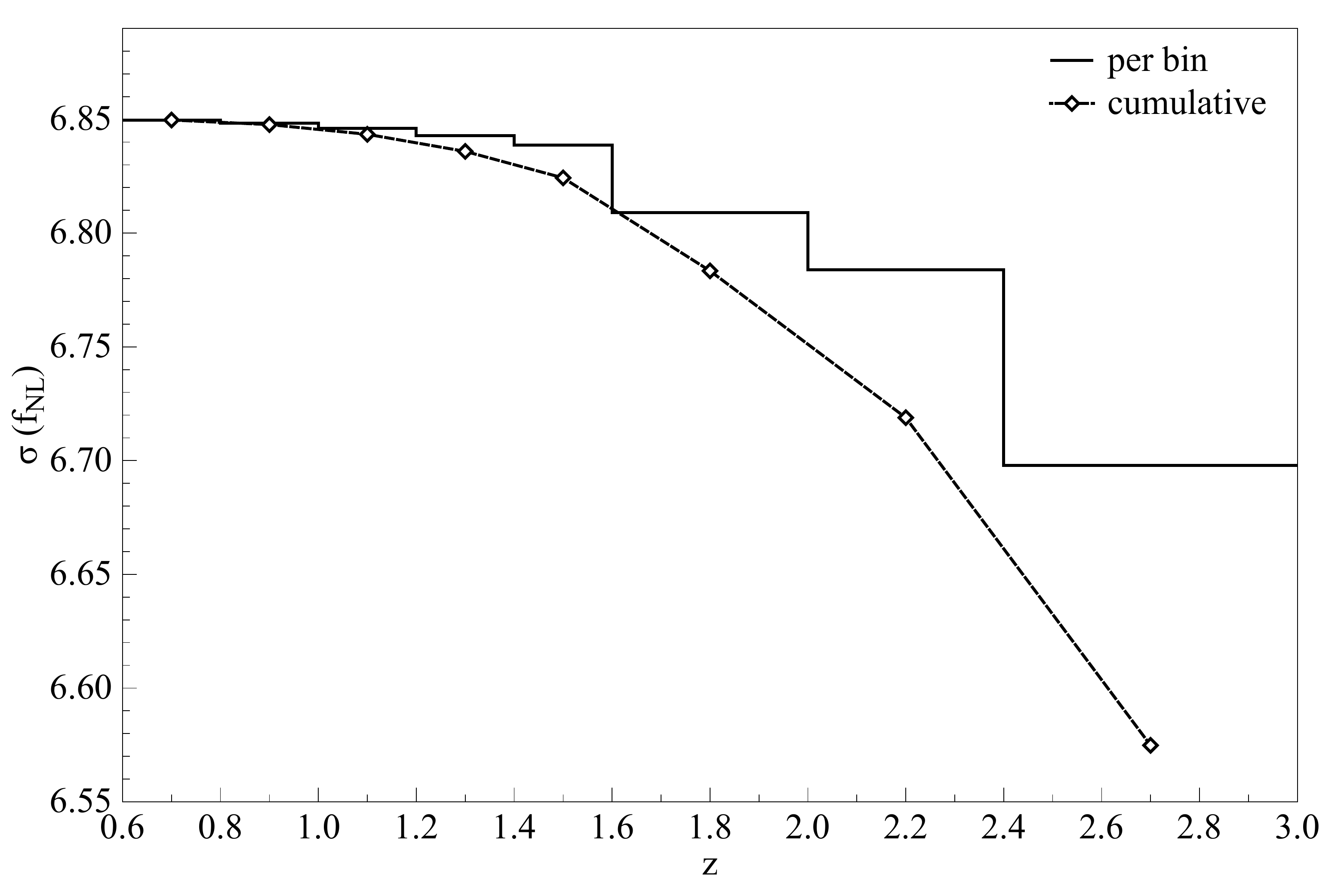}
\includegraphics[width=0.45\columnwidth]{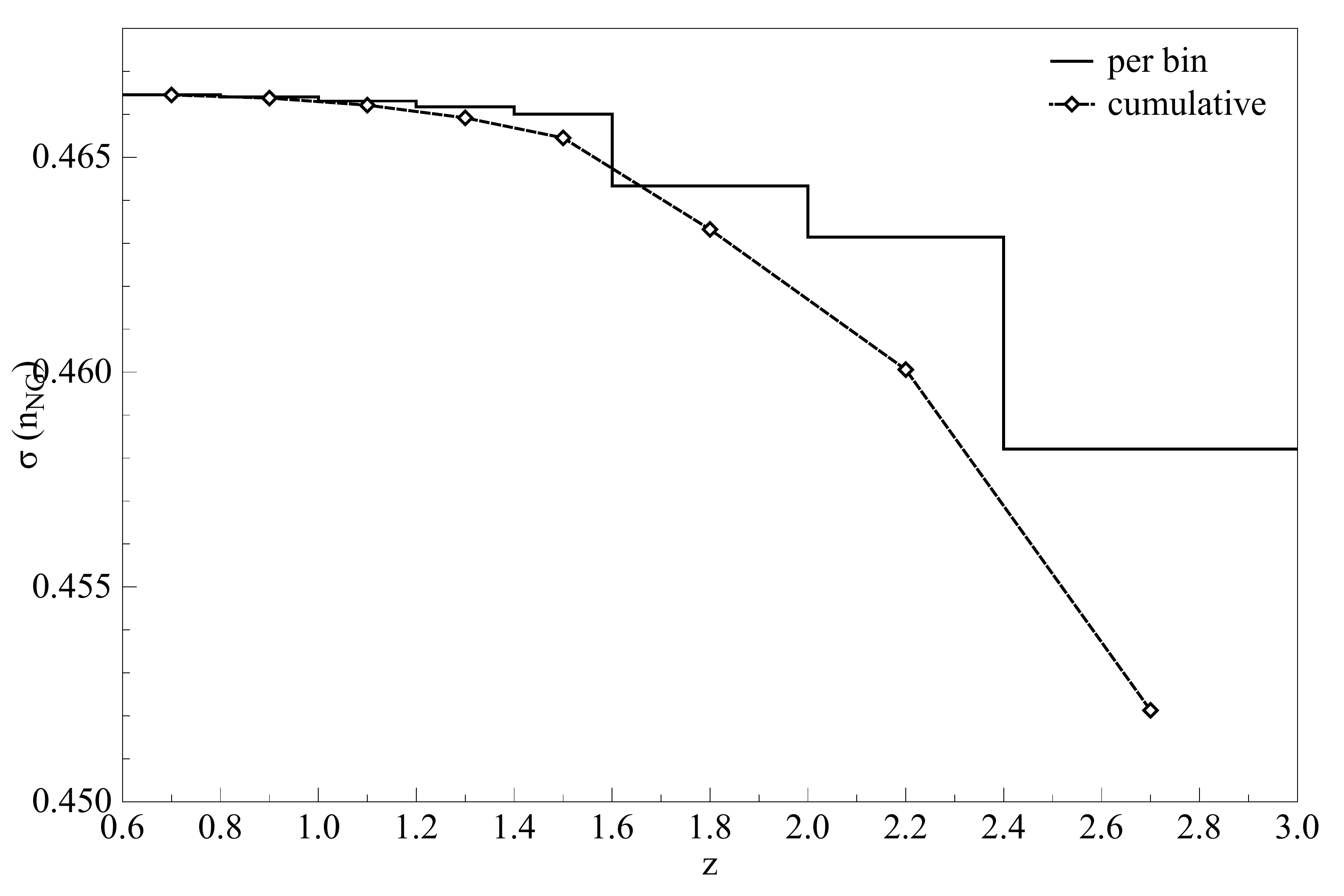}
\caption{Effect of doubling $nP$. The solid line shows the constraining power of individual bins, the dashed line with symbols the cumulative effect of having $nP=2$ for an increasing number of bins.}
\label{fig:fnl_ng}
\end{figure*}
\end{center}

Here we look at the dependence of $\sigma (f_{\rm NL}, n_{\rm NG})$ on the number density.
We start from the reference survey of Table~\ref{tab:sur} and we double $n_g$ in every redshift bin.
The observed number density can either be increased by spending more time surveying a region of the sky, or increasing the number of fibers of the instrument. In Figure~\ref{fig:fnl_ng} we plot the effect of doubling $nP$, per individual bins and its  cumulative effect when doing it for more bins. It can be seen that, after reaching $nP\gtrsim 1$ there is not much to gain when increasing number density (this is true also for dark energy measurements, see e.g.~\cite{seo}).
This is true in the single-tracer case we analyze, while with multiple tracers, the errors are shot noise limited, i.e. there will be a substantial improvement when going to very large number density.

\subsection{Area}
\begin{center}
\begin{figure*}[htb!]
\includegraphics[width=0.45\columnwidth]{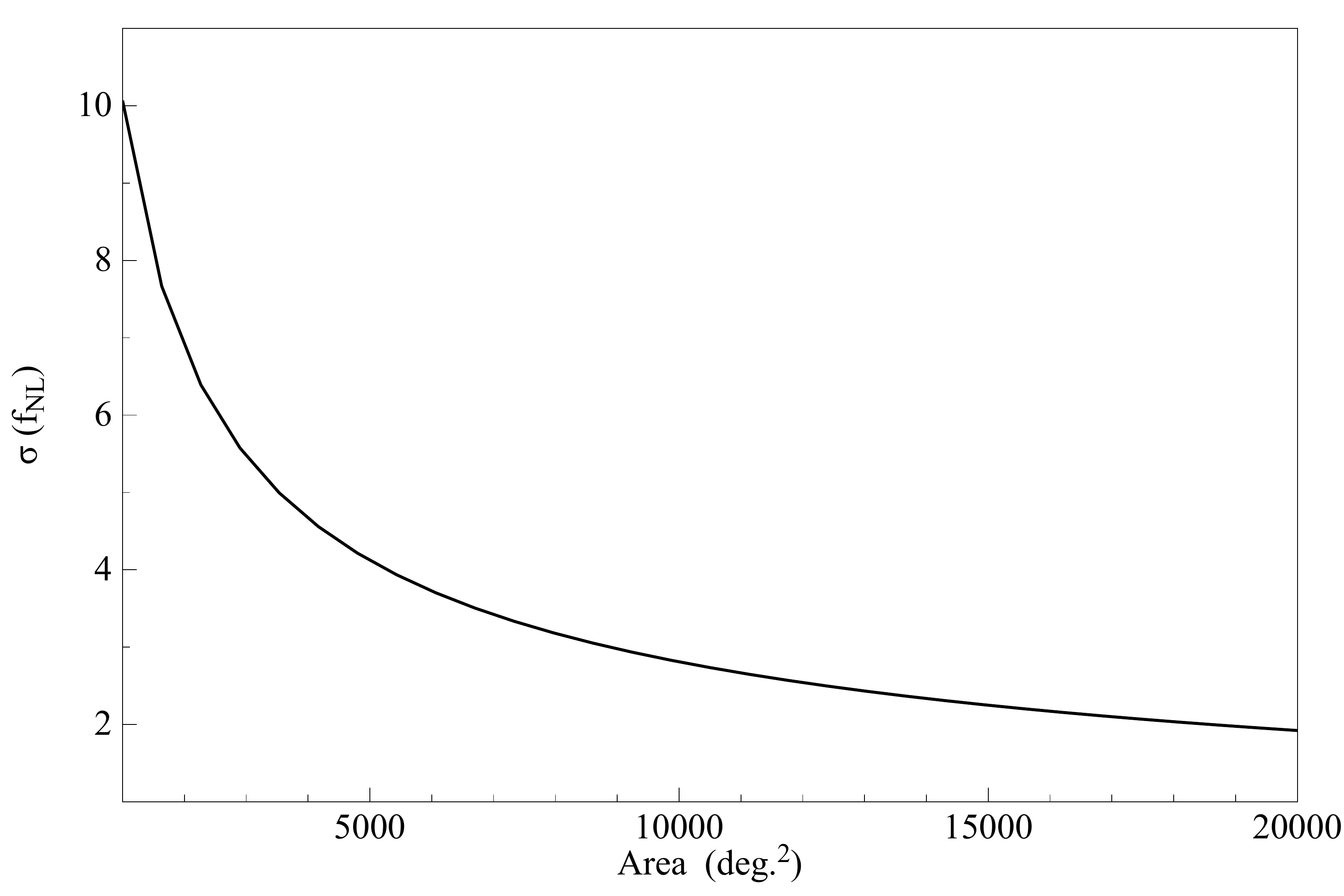}
\includegraphics[width=0.45\columnwidth]{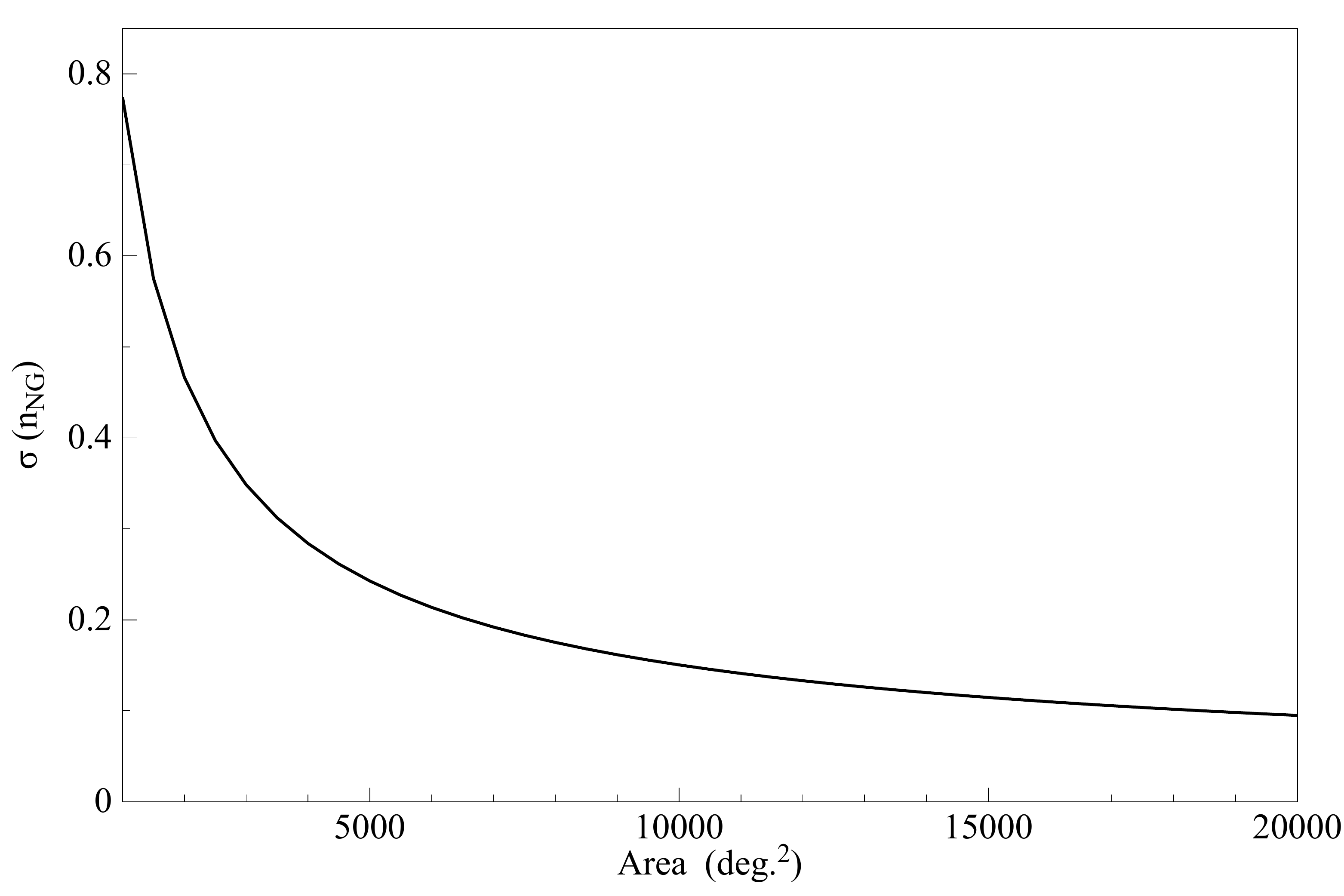}
\caption{Effect of increasing the area surveyed, for a survey with $nP=1$, with the other parameters as in Table~\ref{tab:sur}.}
\label{fig:fnl_area}
\end{figure*}
\end{center}

We now investigate the dependence of $\sigma (f_{\rm NL}, n_{\rm NG})$ on the area surveyed. Larger area can be achieved increasing observing nights or the field of view, or reducing either exposure or overhead time (but maintaining number density). In Figure~\ref{fig:fnl_area} we show the effect on the constraining power for both parameters when increasing the area surveyed, maintaining all the other survey parameters unchanged. We can see that, as expected, increasing the area is much more beneficial than increasing the number density, once $nP=1$ is reached.
However, it should be noted that this would require an increase in the given observing time.

\begin{center}
\begin{figure*}[htb!]
\includegraphics[width=0.45\columnwidth]{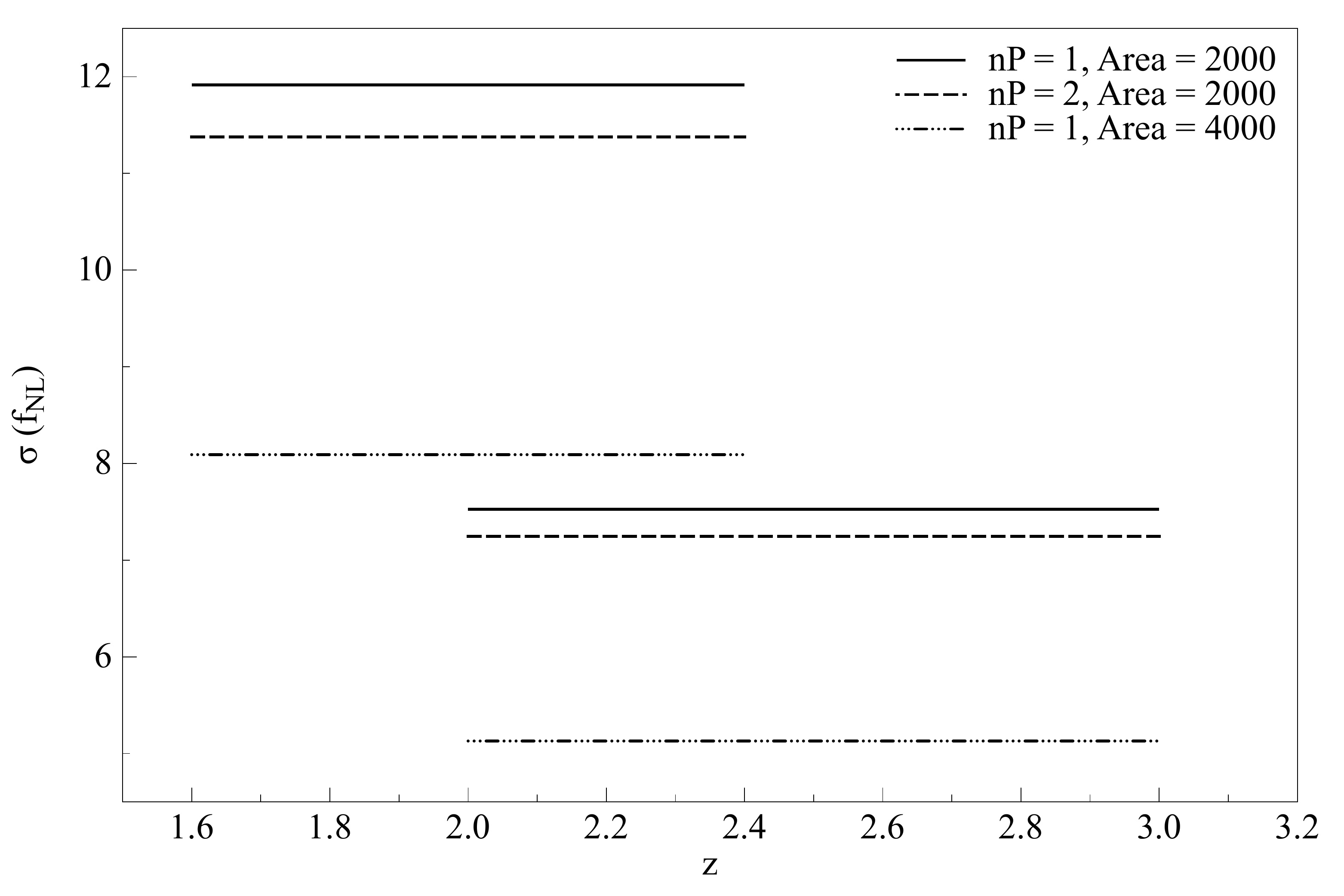}
\includegraphics[width=0.45\columnwidth]{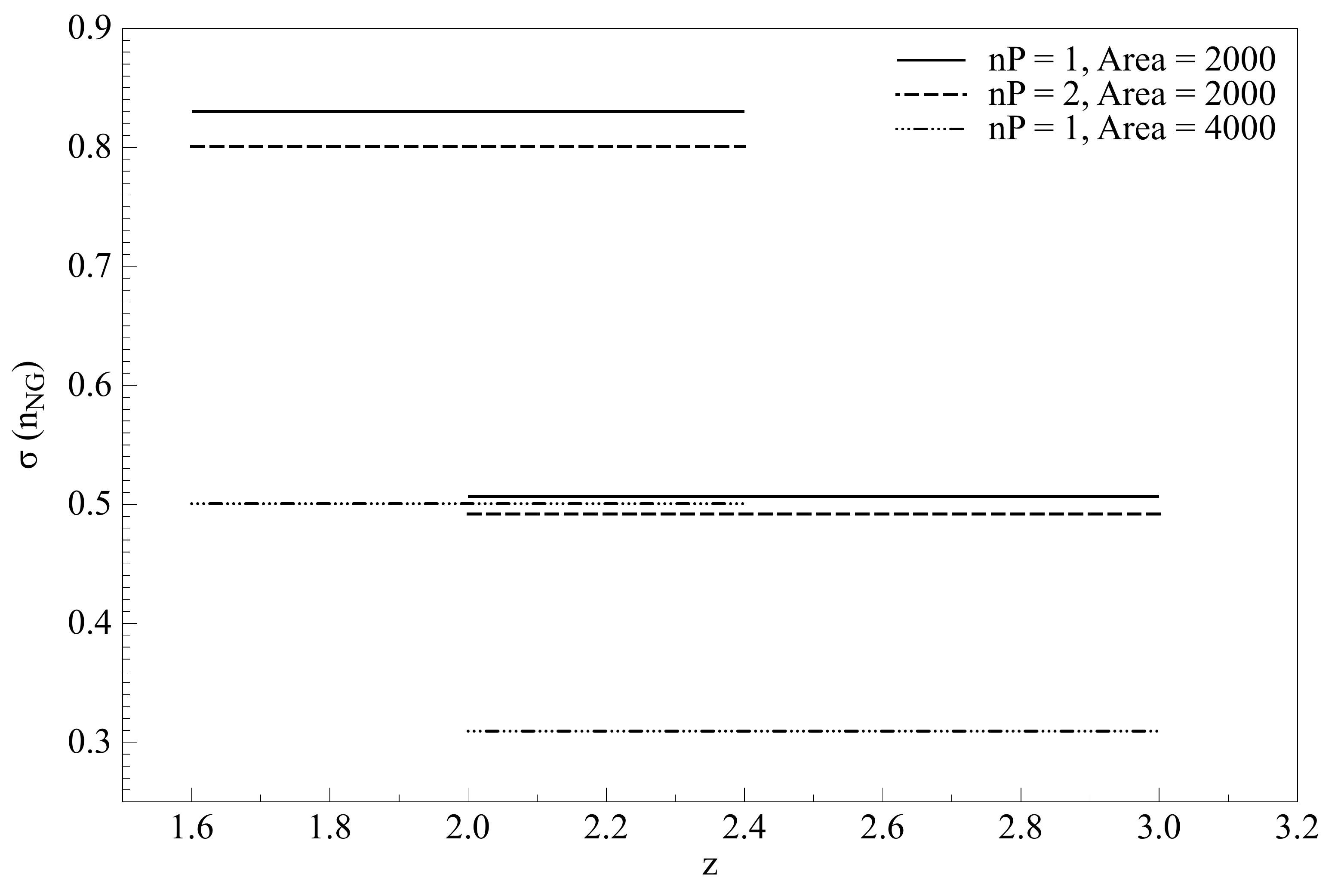}
\caption{High-z constraining power, and comparison of doubling the $nP$ (dashed line) or the area (tri-dotted-dashed line); the dotted line is the constraining power of the fiducial survey over the entire redshift range, and solid lines show the high-z bins with fiducial parameters for \sigfnl ({\it left panel}) and \signng ({\it right panel}).}
\label{fig:fnl_lim}
\end{figure*}
\end{center}

In Figure~\ref{fig:fnl_lim} we plot the precision in the measurement when we use only high redshift bins. We consider two cases: i) $1.6 \le z \le 2.4$, and ii) $2.0 \le z \le 3.0$.
We compare the fiducial case with an area of 2000 deg.$^2$ and $nP=1$ with the effect of doubling the $nP$ or the area. As expected, there is more to gain when increasing the area. 
The scaling of errors showed in this Section is a combination of the number of modes contributing to $f_{\rm NL}$ in the redshift slices and the value and redshift evolution of the number densities and their scale-dependent part of the bias.
It is worth noting that these results are valid for the non-Gaussianity parameters that we are considering here and are not in general valid for other parameters.

\subsection{Populations Targeted}
\begin{center}
\begin{figure*}[htb!]
\includegraphics[width=0.45\columnwidth]{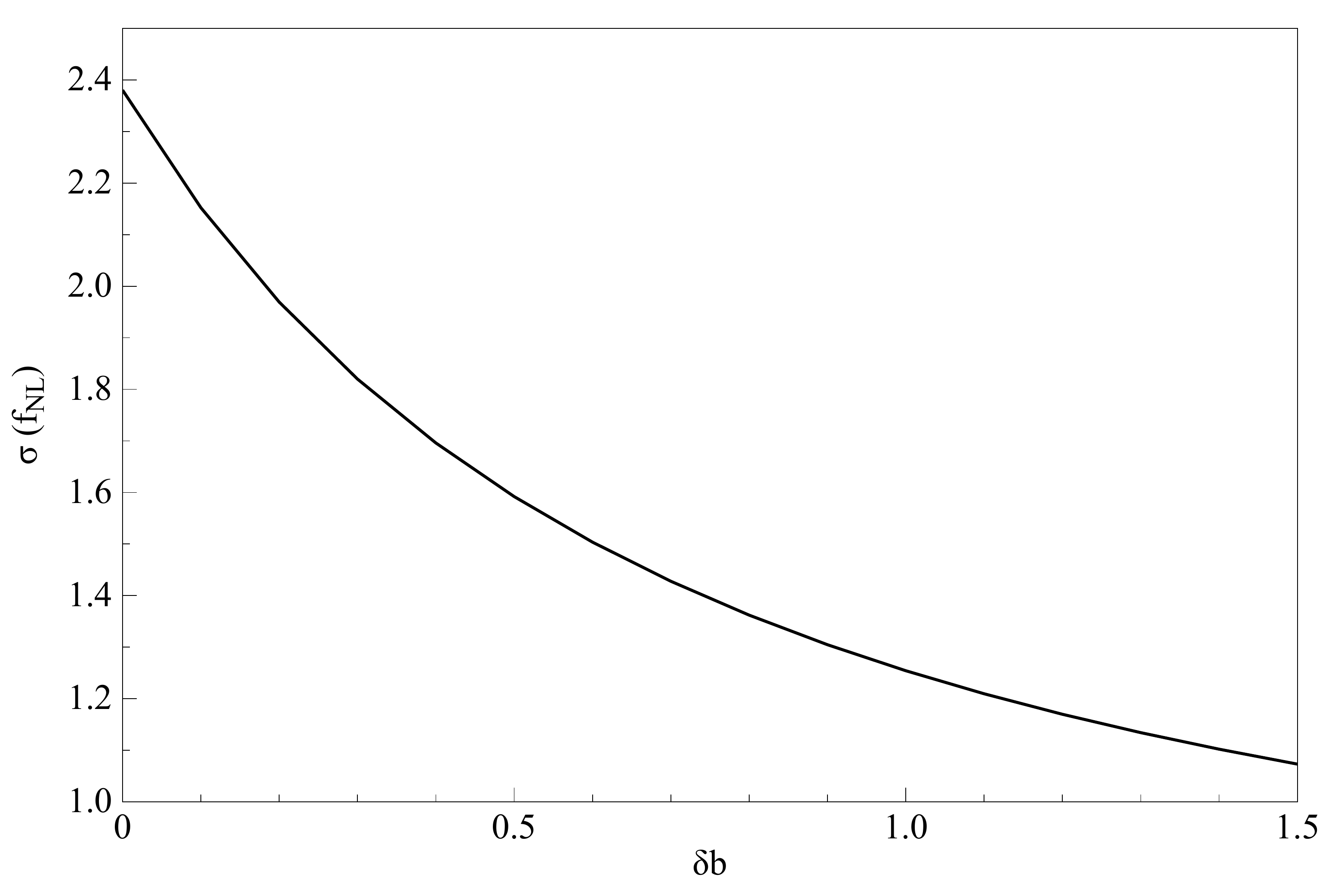}
\includegraphics[width=0.45\columnwidth]{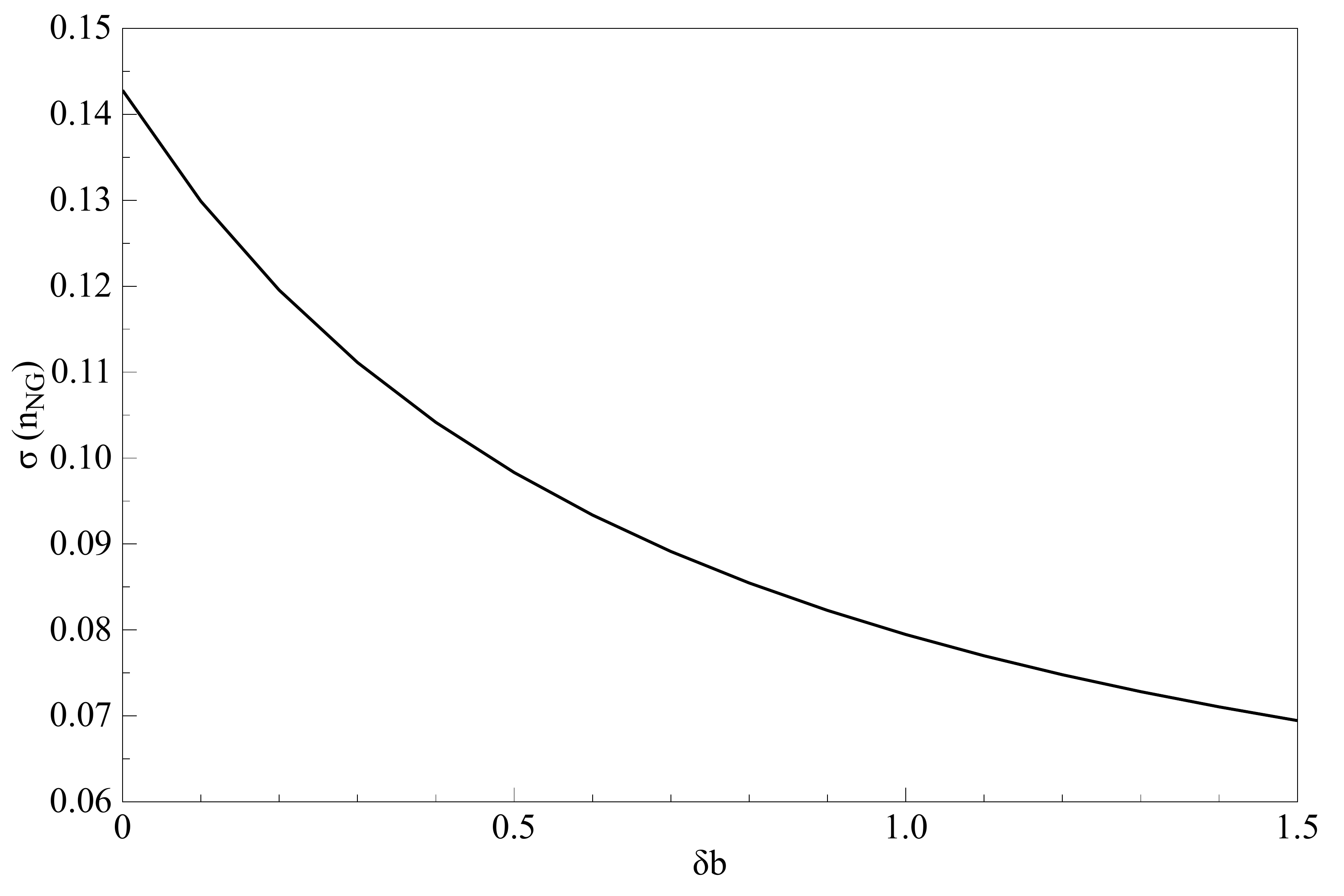}
\caption{Effect of targeting objects with larger bias. On the x-axis is $\delta b$ as defined in the text. Results are computed for the $nP=1$ survey over 20,000 sq. deg., for \sigfnl ({\it left panel}) and \signng ({\it right panel}).}
\label{fig:bias_add}
\end{figure*}
\end{center}

The effect of primordial non-Gaussianity depend on the value of the bias, so here we investigate how the constraints vary when we target objects with a different bias (e.g. LRG). In this paragraph we define the bias as in Equation~\ref{eq:gau-bias}, but with an additive parameter:
\begin{equation}
\label{eq:bias_add}
b_{\rm G}(z) = 0.9 + 0.4 z + \delta b \, . 
\end{equation}
In Figure~\ref{fig:bias_add} we plot the increase in constraining power of primordial non-Gaussianity parameters as a function of $\delta b$.

\begin{center}
\begin{figure*}[htb!]
\includegraphics[width=0.45\columnwidth]{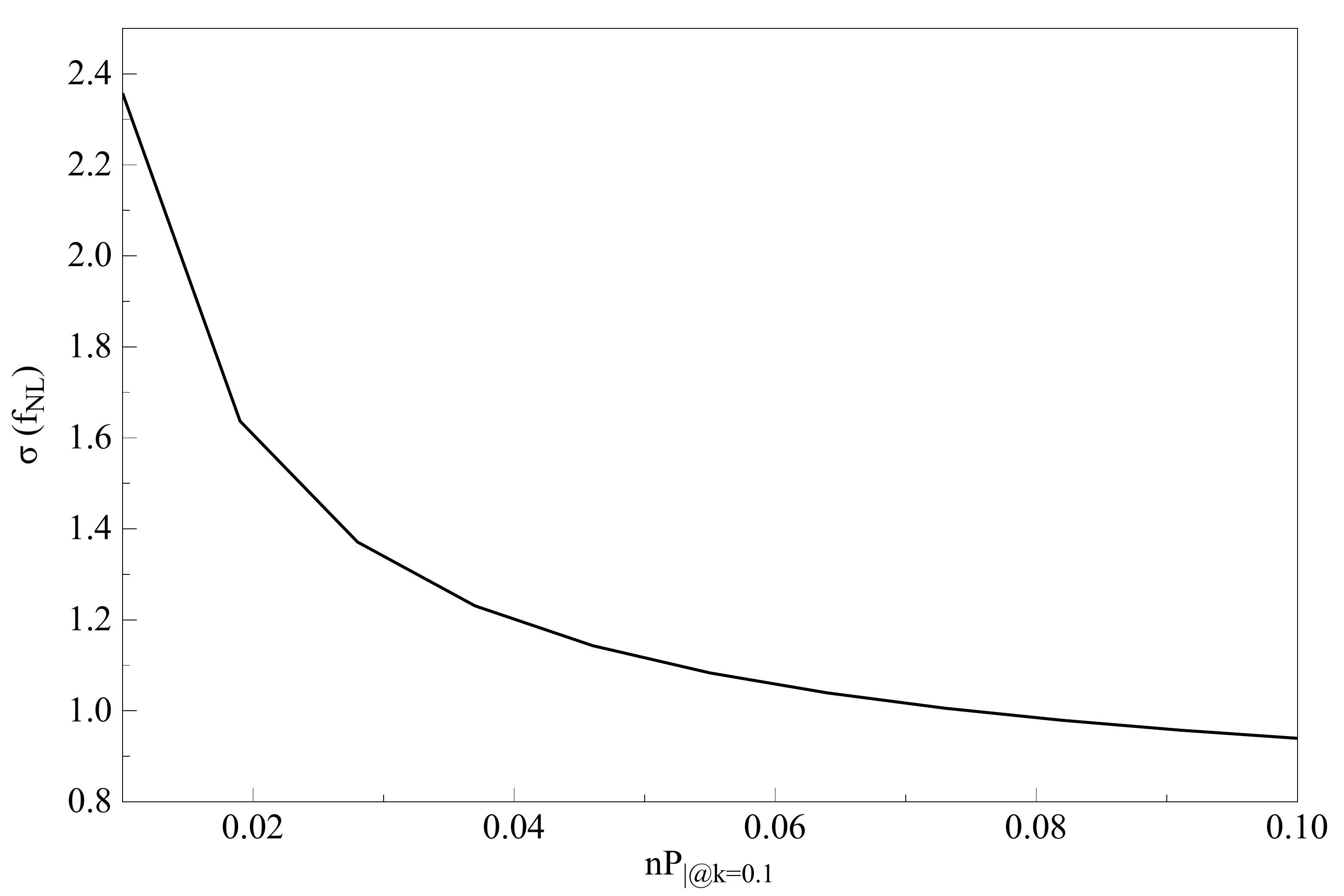}
\includegraphics[width=0.45\columnwidth]{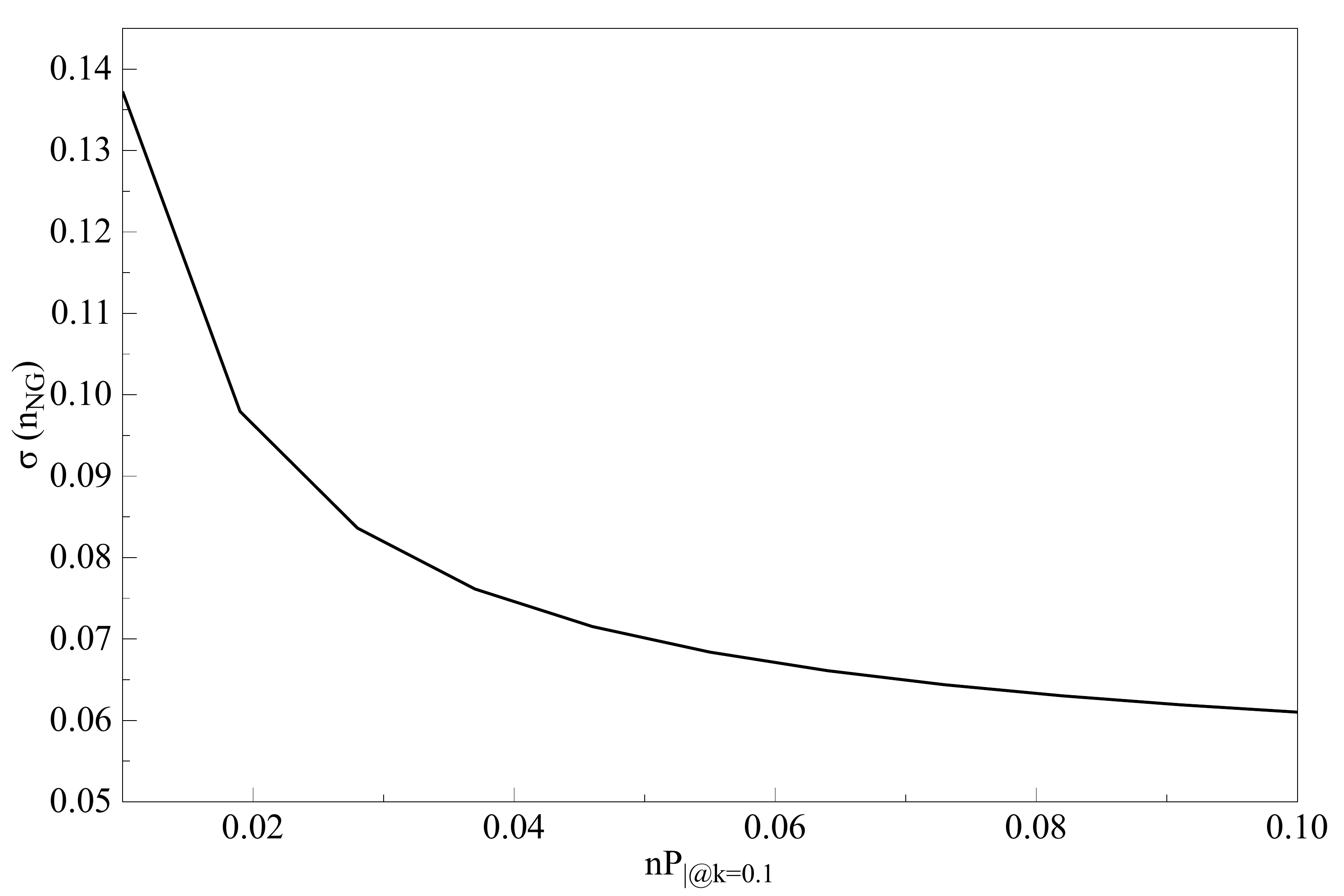}
\caption{Constraining power of a Quasar survey, for \sigfnl ({\it left panel}) and \signng ({\it right panel}).}
\label{fig:qso}
\end{figure*}
\end{center}

To exploit the increased constraining power coming from high-redshift sources and large bias, we also investigate the limits on non-Gaussianity parameters from a quasar survey. We assume a redshift range of $1.8<z<4.0$ in 4 bins, 20,000 sq. deg., and a number density in a range from $0.01$ to $0.1$ of the $nP=1$ survey used for galaxies. In Figure~\ref{fig:qso} we present the errors on \fnlt and \nngt parameters of such configuration, as a function of the number density.

From Figure~\ref{fig:bias_add} and~\ref{fig:qso} we can appreciate how the gaussian bias plays a big role in measuring non-Gaussianity parameters; we can then say that it would be really important to target highly biased objects.

In order to improve the constraining power by reducing the cosmic variance, it has been proposed (see e.g.~\cite{mcdonald}) to use the so-called multiple-tracers technique: one can trace multiple populations and use them as different tracers.
The gain strongly depends on the populations targeted, their relative bias and number density, and the halo mass, and in order to have a sensible increase in the constraining power a very high number density is required.
For detailed analyses of the advantages of this technique, see e.g.~\cite{seljak08, hamaus, yoo12}; the multi-tracer technique has been recently used in real data analyses in~\cite{blakemulti, rossmulti}.
In an ideal scenario one would be able to target a large number of unbiased (so that their shot noise is almost negligible) and a population of strongly biased objects.
In this case the improvement in the error on the \fnlt parameter can be written as~\cite{seljak08} $\left[2/(n_b P_b) \right]^{-1/2}$, where the subscript $_b$ refers to the biased objects.
Recently~\cite{yamauchi14} investigated the improvements given by the multi-tracer technique in the case of a combination SKA-Euclid photometric surveys, while~\cite{ferraro14} studied the ability of a future survey to test multifield inflation.
Their results are in line with our findings and they highlight how the impact of the multi tracer technique depend on the characteristics of the survey.


\section{Survey Optimizations}
\label{sec:opt}
After studying how measurements of non-Gaussianity parameters depend on different survey characteristics, we now investigate what is the optimal configuration in order to have the best constraints possible, given a fixed budget. The precision in \sigfnl depends on $\{A,n_g,b,\hat{z}\}$, where $A$ is the area surveyed, $n_g$ the number density, $b$ the bias and $\hat{z}$ the redshift range. Area and number density depend on the instrument specifications, like the total observation time of the survey $\tau$, the exposure time $t$ (including overhead), the number of fibers, and the field of view.
The detailed relations between these variables and the resulting observed number density are instrument-specific (analyses of that can be found e.g. in~\cite{parkinson, pfs, euclid, bbs}), and in this work we look at what would be needed in order to achieve certain precision in the measurements. Once a requirement is set, it can be achieved in different ways, e.g. one can reach a certain number density by increasing the number of fibers or the exposure time.

\begin{center}
\begin{figure*}[htb!]
\includegraphics[width=0.49\columnwidth]{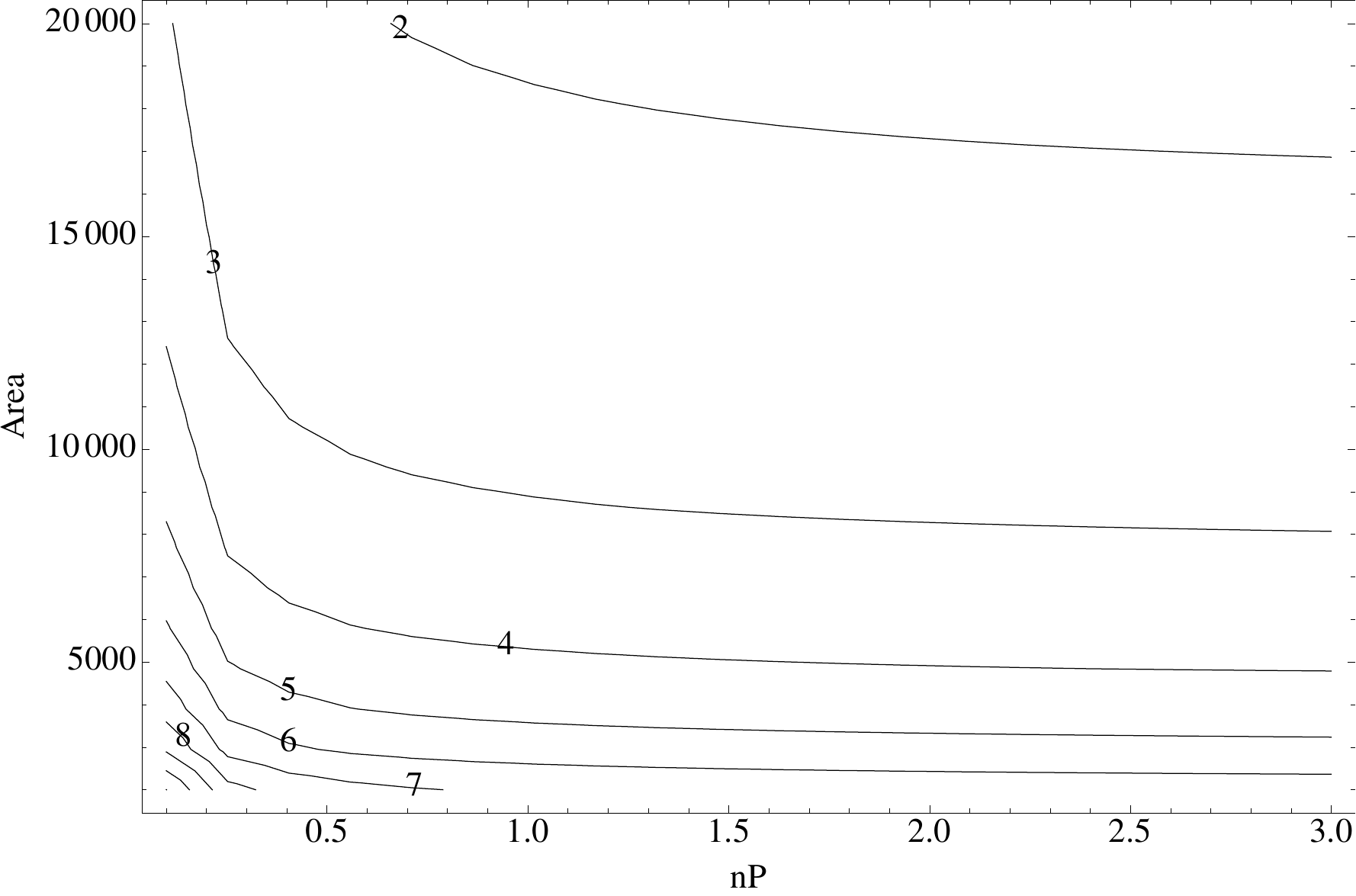}
\includegraphics[width=0.49\columnwidth]{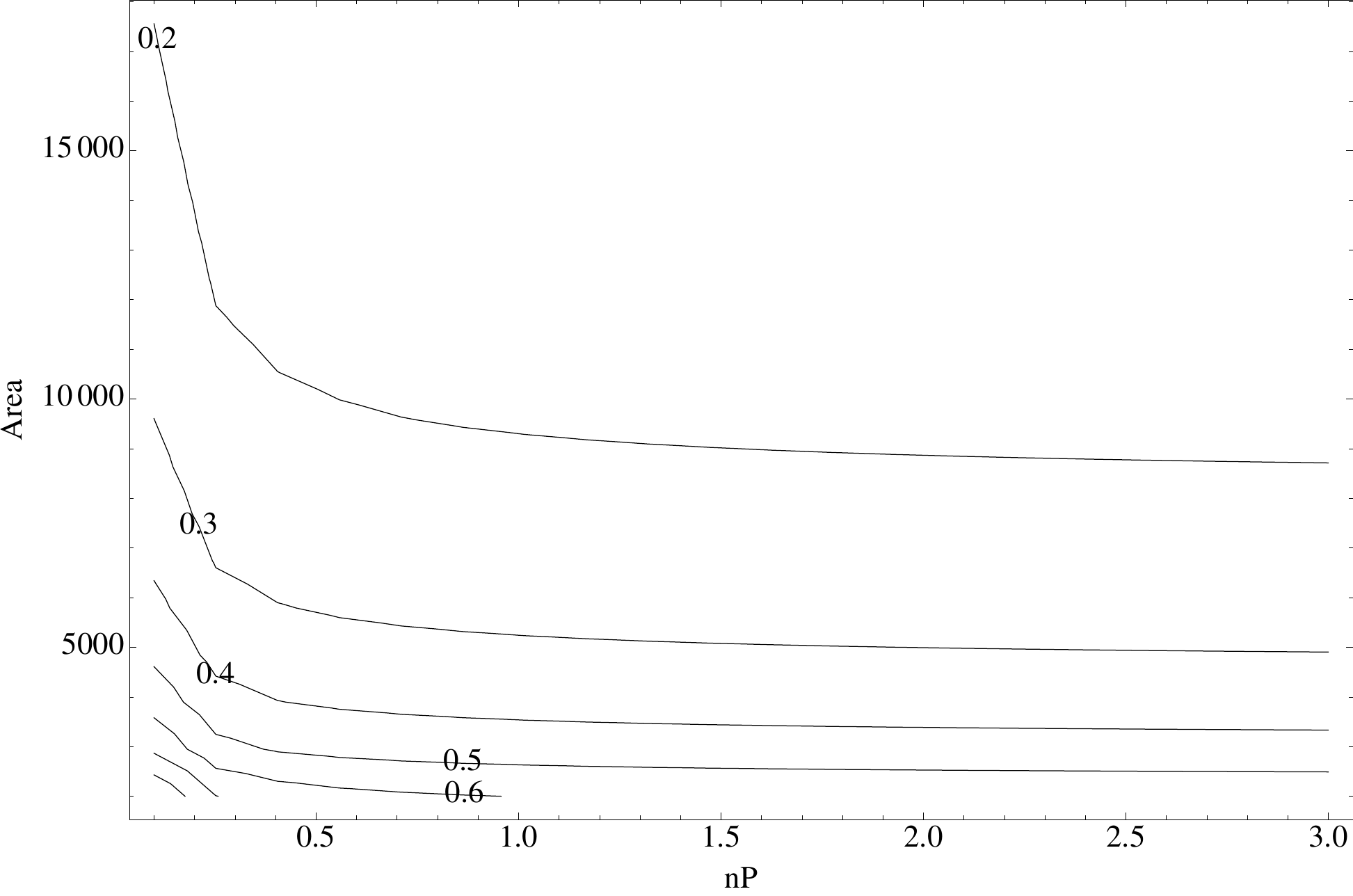}
\caption{Contour plots with precisions in the measurements.
{\it Left Panel}: for $f_{\rm NL}$, as a function of area and $nP$ (as a constant over the entire redshift range).
{\it Right Panel}: for $n_{\rm NG}$, as a function of area and $nP$ (as a constant over the entire redshift range).}
\label{fig:fnl_AnP}
\end{figure*}
\end{center}

In Figure~\ref{fig:fnl_AnP} we show contour plots giving the constraints on \fnlt and \nngt, as a function of area and number density (assumed to be constant over the entire redshift range). It is apparent in both cases that once a threshold of $nP \gtrsim 0.5$ is reached, the gain in constraining power comes from surveying a larger area.
However, for the case of \fnlt, spectroscopic surveys will not be able to reach an uncertainty of $1$ without the multi-tracer technique, and even in that case it is unlikely such a survey could provide a strong detection of $f_{\rm NL} =1$.

In the case of \nngt, it will be relatively easy to bring the uncertainty below unity, but once \signng $\approx 0.2$, then improving the precision will become very difficult.

\subsection{Future Surveys}
For the future surveys considered, we take the baseline survey parameters and scale area and number density, making the rough assumption that they are inversely proportional. This is a realistic assumption when the survey is targeting one type of objects, as in the case of e.g. the PFS survey, where one does not have redshifts for all the objects that could be detected. In this case, the trade-off is between adding new patches of the sky, or spending more time in the same area. When observations get deeper, one starts detecting other types of objects, and in this case, the relation between area and number density for a fixed budget becomes more complicated. For more details, see~\cite{dePutter13}.

\subsection{Combined constraints}
\begin{center}
\begin{figure*}[htb!]
\includegraphics[width=0.47\columnwidth]{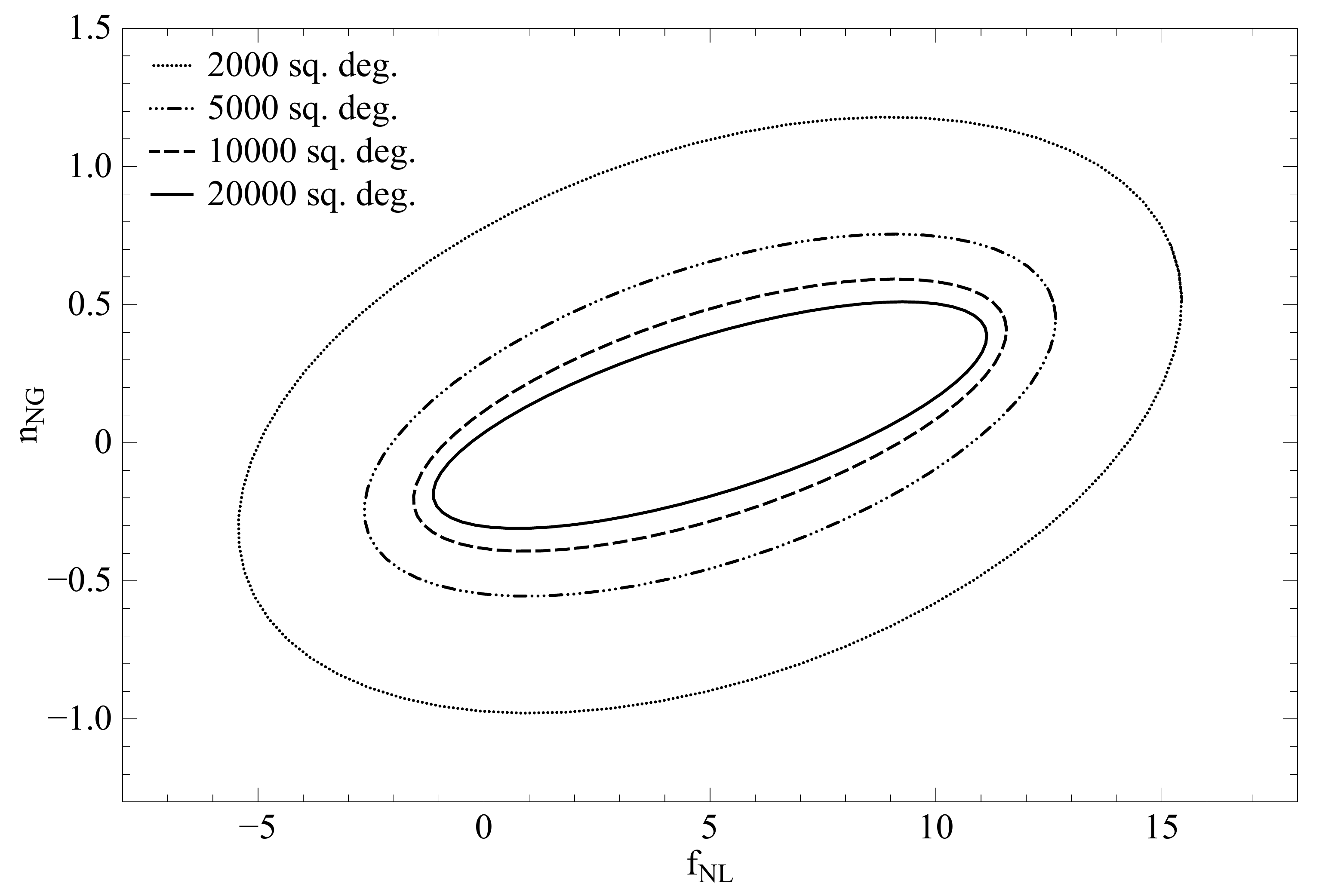}
\includegraphics[width=0.47\columnwidth]{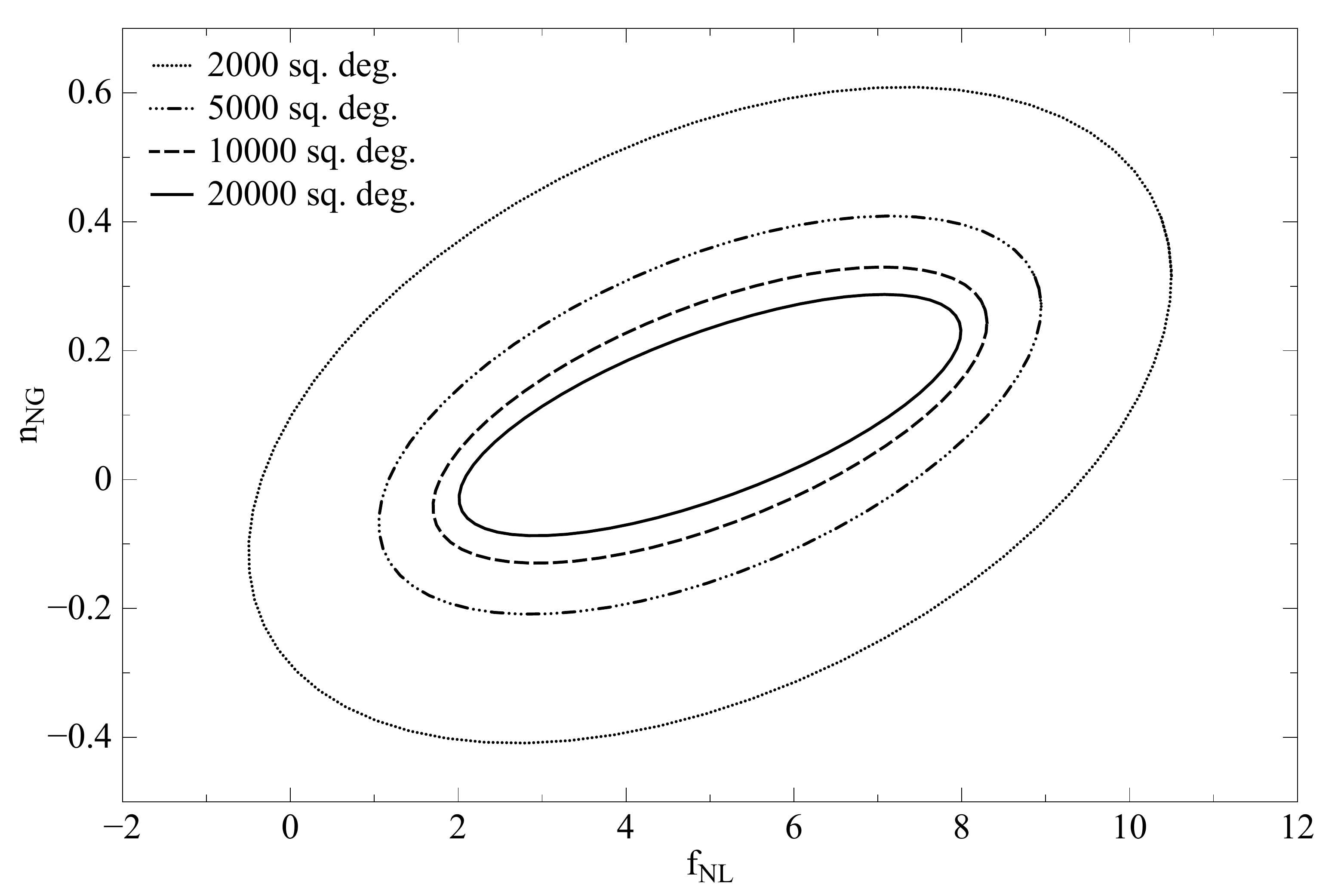}
\caption{Forecasted constraints on the measurement of $\{ f_{\rm NL}, n{\rm NG} \}$ for different configurations of galaxy surveys, when scaling area and number density as described in the text. {\it Left Panel}: for galaxy survey alone; {\it Right Panel}: when adding a Quasar catalog.
}
\label{fig:ell_pfs}
\end{figure*}
\end{center}

In this section we investigate the combined constraints of $\{ f_{\rm NL}, n_{{\rm NG}} \}$, for some configurations of the survey we considered so far. We compute the constraints marginalizing over all the parameters that are not measuring non-Gaussianity (in principle, deviations from the gaussian bias depend also on cosmological parameters and the model of gravity, but here we assume $\Lambda$CDM+GR). Our final Fisher matrix is then a $2 \times 2$ matrix $F_{\alpha \beta}$, with $\{ \alpha, \beta \} = \{ f_{\rm NL}, n_{\rm NG} \}$. In Figure~\ref{fig:ell_pfs} we show the predicted combined constraints on $\{ f_{\rm NL}, n_{\rm NG} \}$ from some configurations of the surveys considered. 
The previous plots show the constraints for the parameters taken singularly in order to see the effect of area, z-range and number density, if e.g. one assumes no running and wants to measure $f_{\rm NL}$ (or knows the value of $f_{\rm NL}$ and wants to focus on its running). The final constraints in the case of non-zero $f_{\rm NL}$ plus running have to be inferred from the plot of Figure~\ref{fig:ell_pfs}, as there is a degeneracy between them.

\section{Conclusions}
\label{sec:conclusions}

We studied trade-offs of specifications for guiding survey design in order to measure non-Gaussianity signatures in the 3D power spectrum with a precision competitive with CMB experiments.
A full survey optimization investigation would require a more detailed modeling of the surveys to be optimized.
This work studies parameter dependencies and should be used as a general guide for planning experiments aiming to measure non-Gaussianity.
When designing a survey, one has to make a choice and decide to either go deeper, have an increased number density, or survey more area. From our calculations, we found that once the number density is somewhat around a value of $nP|_{k=0.1}=1$, adding more sources does not effectively improve the constraining power. On the other hand, increasing the volume helps but not after the area is roughly half the sky; at that point, going to higher redshift will improve the precision in the measurements of \fnlt and $n_{\rm NG}$.

Our results show that forthcoming spectroscopic surveys could give an error on the estimate on the non-Gaussianity parameter $\sigma (f_{\rm NL}) \approx 3$, value that could be decreased to $\approx 1$ when including a quasar population at high redshift and with large bias, when assuming no running of $f_{\rm NL}$.
In the case of $f_{\rm NL}$ plus running, it will be very difficult to bring the uncertainty on $f_{\rm NL}$ below $\approx 5$, while for the running parameter $\sigma (n_{\rm NG})$ the best result will be to have an uncertainty of $\approx 0.3$.
Even when adding a high-redshift quasar population, the combined constraints on $\{ f_{\rm NL}, n_{\rm NG} \}$ are $\approx \{ 3, 0.1 \}$.
For the $f_{\rm NL}$ parameter, it is expected to have a detection when reaching a value of $1$; our results show that for galaxy surveys to reach that target, it will be required to observe a high number density of objects (so to take advantage of the multiple tracer technique) and a large area surveyed.

However, while this result can be competitive in constraining non-Gaussianity parameters, some other measurements will probably be required to reach the precision required to adequately test the primordial universe, such as higher order statistics of the large-scale structure of the Universe. 
It is also worth noting that radial modes on very large scales (see e.g.\cite{Raccanelliradial}) can in principle improve constraints on non-Gaussianity parameters, so the results here presented might be pessimistic.
A more detailed investigation of the constraining power of these measurements will be presented in a future work.

\vspace{0.5 cm}

{\bf Acknowledgments:}\\
Part of the research described in this paper was carried out at the Jet Propulsion Laboratory, California Institute of Technology, under a contract with the National Aeronautics and Space Administration.
We would like to thank Dragan Huterer and Vincent Desjacques for useful suggestions,
and David Bacon, Nicola Bartolo, Sabino Matarrese and Will Percival for useful discussions.

\end{document}